# Molecular detection and antimicrobial activity of Endophytic fungi isolated from a medical plant *Rosmarinus officinalis*

**Shimal Yonuis Abdulhadi, Ghazwan Qasim Hasan\* and Raghad Nawaf Gergees**

Department of Biology, College of Education for pure science, Mosul University, Mosul, Iraq.
\*For correspondence. E-mail: ghazwan.qasim@yahoo.com.

**ABSTRACT**

Endophytes are tiny organisms present in living tissues of distinct plants and have been extensively studied for their endophytic microbial complement. Roots of Rosmarinus officinalis were subjected to the isolation of endophytic fungi and screened for antimicrobial activity against Gram-positive (*Staphylococcus aureus* and *Bacillus subtilis*) and Gram-negative (*Escherichia coli, Pseudomonas aeruginosa, Klebsiella pneumoniae*) bacteria. Genomic DNA from active fungal strain of *Trichoderma harzianum* was isolated, and the internal transcribed spacer (ITS) region was amplified using ITS4 and ITS5 primers and sequenced for genetic inference in fungus. The crude extract of *T. harzianum* isolate with Ethyl acetate was showed significant antimicrobial activity against *P. aeruginosa, S. aureus, K. pneumonia, B. subtilis* and *E. coli*. The antimicrobial activity was highest against *P. aeruginosa* at concentration of 40μg/ml, followed by S. aureus and K. pneumonia at the same concentration. The lowest antimicrobial activity was against by S. aureus at concentration of 60μg/ml. The current study is confirmed that the antimicrobial activity is due to bioactive compounds founded in endophytic fungi.

## INTRODUCTION

*Rosmarinus officinalis*, commonly known as Rosemary, is a member of mint family Lamiaceae. The name "rosemary" derives from Latin ros marinus ("dew of the sea") (Room, 1986). Plant roots are not only colonized by mycorrhizal bacteria and rhizobial, but also by endophytic fungi (Girlanda and Luppi, 2006). The endophytic fungi, do not cause obvious disease, but colonize different plant tissues. For example, can improve host growth, provide the host with some nutrients, convey stress tolerance, and many other benefits. In contrast to endophytic growth in the ground plant organs, growing within the roots has frequently been found to be extensive (Stone et al., 2000, Schulz and Boyle, 2005). The interaction between the plant and endophyte be based on the disposition of host and fungus and the environmental conditions, this relationship could be neutral, mutualistic or antagonistic. Some endophytic fungi may adopt mycorrhizal functions and gives the plants opportunity to competitive other organisms such as insect pests, herbivores, or bacteria (Carroll, 1988; Hawksworth, 1991). Other endophytes can be pathogenic to the host when conditions are unsuitable (Schulz et al. 1999). The biodiversity of root endophytic communities differs depending on the type of vegetation , environmental factors and the interactions between microorganisms. About 80% of endophytic fungi produce secondary metabolites that active biologically in vitro and have fungicidal, antibacterial and herbicidal activities (Schulzetal.2002). For instance, *Fusarium* spp. colonize both the roots and shoots in many hosts, and synthesis a number of toxins including beauvericin, a cyclic hexadepsipeptide to protect infected plants against herbivory insects (Kuldau and Yates 2000; Miller, 2001). Moreover, fungal root endophytes produce metabolites toxic to the nematode *Meloido gyneincognita.* For instance, both phomalactone are produced by *Verticillium chlamydosporium* (Khambay et al., 2000), and secondary compounds present in the culture filtrate of *F. oxysporum* decreased the mobility of the nematode within10 minutes after exposure (Hallmann and Sikora, 1996).

In addition, several root endophytic fungi synthesis antimicrobial metabolites (Schulz et al., 2002), e.g. the antibacterial and antifungal metabolites synthesized in vitro by *Cryptosporiopsis* sp. from *Larix decidua* (Schulz et al., 1995), which help to protect the host from other organisms. In addition, the endophytic *Neotyphodium* spp. can be stimulated in shoots of the host after the damage, to increase production of mycotoxins as a kind of adaptive (Bultman and Murphy, 2000); the same effect can be occurred in the injured roots which colonized by endophytic fungi. Furthermore, the fungal roots can synthesis biologically active secondary compounds in both in vitro and in vivo (in planta), which able to be antagonistic against other organisms such bacteria, and lead to suppress the disease; it could also play a vital role in keeping a balance of antagonism between endophyte and host.

In the other side, many endophytic fungi have shown antimicrobial activity against the human pathogenic bacteria which the current paper about. For instance, the isolated endophytic fungi (*Phomopsis sp.*) from *Vitex negundo* L leaves with ethyl acetate showed significant antimicrobial activity against, *S. typhimurium*, *E. coli*, *B. subtilis*, *K. pneumoniae*, *S. aureus*, and *B. cereus* (Desale and Bodhankar, 2013). The ethyl acetate extract of endophytic fungi *Colletotrichum gloeosporioides* showed potential inhibition against *E. coli, S. aureus*, *S. typhimurium, B. cereus* and *P. aeruginosa*; again the ethyl acetate extract of *Fusarium oxysporum* has the same effect against previous bacteria except *P. aeruginosa* (Ramesha and Srinivas, 2014). Endophyte fungi strains isolated from the roots of *Sonneratia griffithii* Kurz with ethyl acetate and butanol have shown higher antibacterial activity against *S. aureus* and *E. coli* comparing with other parts (Handayani et al., 2017). Endophyte fungi isolated from leaves of Bush mango showed antibacterial activity against some pathogenic bacteria such as *S. aureus*, *B. subtilis*, *E. coli* and *P. aeruginosa* (Nwakanma et al., 2016). The ethyl acetate extracts of mycelia and filtrates isolated from *Vaccinium dunalianum* var. *Urophyllum* revealed a significant inhibit of the growth of pathogenic bacteria such as *B. subtilis*, *S. aureus*, *P. vulgaris*, *L. monocytogenes* and *Salmonella bacteria* (Tong et al., 2018).

## MATERIALS AND METHODS

**Plant samples collection:**
Samples of *R. officinalis* plants were collected from Mosul city, north of Iraq during a week in September 2019. The roots were excised from vegetable parts using a sterile knife and brought to the laboratory in sterile polythene bags.

**Isolation of endophytic fungi:**
The roots were washed with water several times to remove the soil and the impurities sticking to it. The roots were cut into small pieces (5 mm × 2 mm) using sterile blade and washed with sterile distilled water thrice and allowed to surface dry under aseptic conditions (Nithya and Muthumary, 2010). Then samples were dipping into 96% ethanol for one minute and then in 2% sodium hypochlorite for three minutes and 70% ethanol for one minute, then washed with sterile water twice and allowed to surface dry by placing them into sterile filter papers. The samples were placed on Potato Dextrose Agar (PDA) plates with adding 50 mg/L tetracycline to inhibit the bacterial growth and incubated at 28ºC for 14 days. The hyphal tip of growing out from root tissue was conveyed to fresh PDA plates amended with 50 mg/L tetracycline. The surface sterilization process was confirmed for efficacy by taking 0.5 ml of washing water from the last step and spreading it on the Petri dishes containing PDA medium, the plates were incubated in the same steps above. The absence of any fungal colony is an indication of the efficiency of the surface-sterilization process.

**Identification of endophytic fungi:**
**Morphological identification:**
Several fungal isolates were appeared. The most common fungi selected was *T. harzianum* as shown in Figure1.

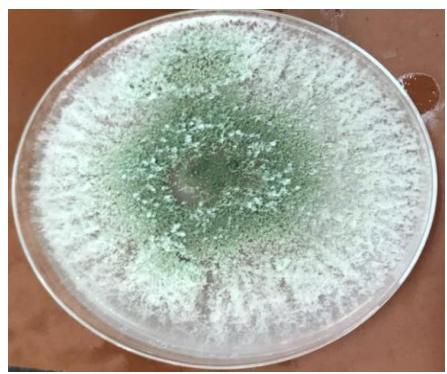

Figure1: Culture plate of *T. harzianum* which was marked by the green color.

**DNA extraction, PCR amplification and molecular detection:**

The *T. harzianum* were grown in 100 ml Potato Dextrose Broth (PDB) for 10 days at 25º C. The Mycelia were collected from the PDB and washed with distilled water and ground using sterile pestle and mortar with liquid nitrogen until dry powder was obtained. 1.0 g DNA of fresh mycelium was extracted (Saghai Maroof et al., 1984). The nucleic acid was extracted using the cetyl Trimethyl Ammonium Bromide (cTAB) method (O'Donnell et al., 1997). The internal transcribed spacer (ITS) region was amplified by polymerase chain reaction using the universal primers ITS4 (5′- TCCTCCGCTTATTGATATGC-3′) and ITS5 (5′-GGAAG TAAAAGTCGTAACAAGG-3′) (Shweta et al., 2010). The 590- bp DNA fragment PCR products was checked by electrophoresis on a 1 % agarose with ethidium bromide and visualized using a UV transilluminator. The fragment was eluted and purified using and the Agarose Gel DNA Extraction Kit (Takara, Japan). Sequencing of ITS product was sent to macrogen company/ South Korea https://dna.macrogen.com/#. A basic local alignment search tool (BLAST) analysis was carried out in the NCBI database, the results showed a 100% matches between the current sequences and deposited sequences in Genebank (Accession: MH363810.1).

**Fungal Isolates and Culturing Conditions:**
The *T. harzianum* isolates were obtained from *Rosmarinus officinalis* roots growing in Mosul city north of Iraq. For cultivation of fungi, five mycelia plugs taken from actively growing colony margin using cork borer No. 2 (5- mm diameter) were inoculated into a 500 ml conical flasks containing 100 ml PDB and incubated at 25°C for 14 days.

**Extraction of secondary metabolites:**
After incubation the culture filtrate was separated from the fermentation medium using a rotary vacuum evaporator, and filtered through a filter paper type Whaitman No.1, then the pH was adjusted to 3 by adding some drops of 0.1 N HCL. In order to extract the bioactive compounds, the equal amount of ethyl acetate was added to the culture filtrate and allowed to stand for one day using Separating funnel. The organic phase was collected in petri dishes and was evaporated to dryness at room temperature. The dry extract was collected in sterile and tightly closed flasks and put in the refrigerator until use.

**Antimicrobial activity by agar well diffusion method**
Endophytic fungi isolated from a medical plant of *R. officinalis* was subjected to screening for antimicrobial activity against the human pathogenic bacteria *Pseudomonas aeruginosa*, *Staphylococcus aureus*, *Klebsiella pneumonia*, *Bacillus subtilis,* and *Escherichia coli*. The crude extract of *T. harzianum* was dissolved in Dimethyl sulfoxide (DMSO) at different concentrations of 20 μg/ml, 40 μg/ml and 60 μg/ml ; poured into the 5-mm diameter well made in Petri dishes containing 2% nutrient agar (Peptone-5 g, Beef Extract-3 g, NaCl-5 g, distilled water-1000 ml, Agar powder-20 g, pH 7.0). The cultures were kept for 24 hours at 28 °C for the antimicrobial metabolite diffusion and thereafter they were incubated at an appropriate temperature for the growth of test-microorganisms. The zone of inhibition was measured in mm (Visalakchi and Muthumary, 2009). Furthermore, Gentamicin (40 μg/mL) and used as positive control, while Distilled water was used as the negative control (Table 1).

**Results**

**DNA extraction and ITS sequence analysis**
The DNA was extracted using cTAB method, the ITS region was amplified by PCR using the universal primers ITS4 and ITS5. The product was examined by electrophoresis in 1% agarose gels (Figure 2). The purified fragment was sent for sequencing. The sequence data of the ITS region was compared with deposited sequences in NCBI (Accession: MH363810.1) using (BLAST) analysis. The studied ITS sequence result showed a homology of 100% with deposited sequences (Figure 3).

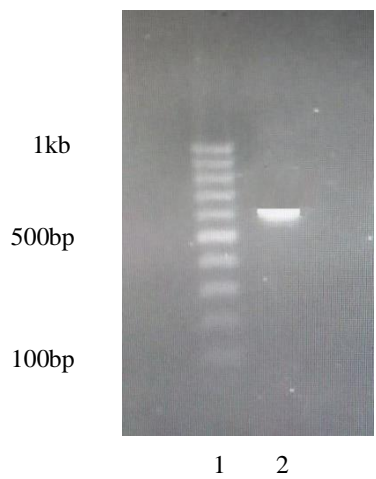

Figure 2: Amplified ITS fragments of *T. harzianum*. Agarose gel image showing: Lane (1) 100 bp DNA ladder, lane (2) ITS PCR product amplified with ITS4 forward and ITS5 reverse primers, Fragment size around 650 bp.

```
Score            Expect    Identities      Gaps        Strand
1090 bits(590)   0.0       590/590(100%)   0/590(0%)   Plus/Plus

Query  1    CAAGATAGCCGTATGAAGGAGGGATATTACCGAGTTTACAACTCCCAAACCCAATGTGAA  60
            ||||||||||||||||||||||||||||||||||||||||||||||||||||||||||||
Sbjct  1    CAAGATAGCCGTATGAAGGAGGGATATTACCGAGTTTACAACTCCCAAACCCAATGTGAA  60

Query  61   CGTTACCAAACTGTTGCCTCGGCGGGATCTCTGCCCCGGGTGCGTCGCAGCCCCGGACCA  120
            ||||||||||||||||||||||||||||||||||||||||||||||||||||||||||||
Sbjct  61   CGTTACCAAACTGTTGCCTCGGCGGGATCTCTGCCCCGGGTGCGTCGCAGCCCCGGACCA  120

Query  121  AGGCGCCCGCCGGAGGACCAACCAAAACTCTTTTTGTATACCCCCTCGCGGGTTTTTTAT  180
            ||||||||||||||||||||||||||||||||||||||||||||||||||||||||||||
Sbjct  121  AGGCGCCCGCCGGAGGACCAACCAAAACTCTTTTTGTATACCCCCTCGCGGGTTTTTTAT  180

Query  181  AATCTGAGCCTTCTCGGCGCCTCTCGTAGGCGTTTCGAAAATGAATCAAAACTTTCAACA  240
            ||||||||||||||||||||||||||||||||||||||||||||||||||||||||||||
Sbjct  181  AATCTGAGCCTTCTCGGCGCCTCTCGTAGGCGTTTCGAAAATGAATCAAAACTTTCAACA  240

Query  241  ACGGATCTCTTGGTTCTGGCATCGATGAAGAACGCAGCGAAATGCGATAAGTAATGTGAA  300
            ||||||||||||||||||||||||||||||||||||||||||||||||||||||||||||
Sbjct  241  ACGGATCTCTTGGTTCTGGCATCGATGAAGAACGCAGCGAAATGCGATAAGTAATGTGAA  300

Query  301  TTGCAGAATTCAGTGAATCATCGAATCTTTGAACGCACATTGCGCCCGCCAGTATTCTGG  360
            ||||||||||||||||||||||||||||||||||||||||||||||||||||||||||||
Sbjct  301  TTGCAGAATTCAGTGAATCATCGAATCTTTGAACGCACATTGCGCCCGCCAGTATTCTGG  360

Query  361  CGGGCATGCCTGTCCGAGCGTCATTTCAACCCTCGAACCCCTCCGGGGGGTCGGCGTTGG  420
            ||||||||||||||||||||||||||||||||||||||||||||||||||||||||||||
Sbjct  361  CGGGCATGCCTGTCCGAGCGTCATTTCAACCCTCGAACCCCTCCGGGGGGTCGGCGTTGG  420

Query  421  GGATCGGCCCTGCCTCTTGGCGGTGGCCGTCTCCGAAATACAGTGGCGGTCTCGCCGCAG  480
            ||||||||||||||||||||||||||||||||||||||||||||||||||||||||||||
Sbjct  421  GGATCGGCCCTGCCTCTTGGCGGTGGCCGTCTCCGAAATACAGTGGCGGTCTCGCCGCAG  480

Query  481  CCTCTCCTGCGCAGTAGTTTGCACACTCGCATCGGGAGCGCGGCGCGTCCACAGCCGTTA  540
            ||||||||||||||||||||||||||||||||||||||||||||||||||||||||||||
Sbjct  481  CCTCTCCTGCGCAGTAGTTTGCACACTCGCATCGGGAGCGCGGCGCGTCCACAGCCGTTA  540

Query  541  AACACCCAACTTCTGAAATGTGACCTCGGATCAGGTAGATACCCCTCAAC  590
            ||||||||||||||||||||||||||||||||||||||||||||||||||
Sbjct  541  AACACCCAACTTCTGAAATGTGACCTCGGATCAGGTAGATACCCCTCAAC  590
```

Figure 3: Multiple alignments of ITS sequences of *T. harzianum* with deposited sequences in Genebank (Accession: MH363810.1

**HPLC analysis**
1 ml of culture filtrate of *T. harzianum* extracted by ethyl acetate was used for high-performance liquid chromatography analysis according to (Mradu et al., 2012), the process of diagnosing phenolic compounds was carried out in the laboratories of the Ministry of Science and Technology / Environment and Water Department after acid hydrolysis process by HPLC model (SYKAM) Germany. Column is 18-ODS and dimensions 4.6 mm* 25 cm. The mobile phase is (Methanol: D.W. : acetic acid) (85:13:2), samples were detected at a wavelength 360 nm at a flow rate of 1 ml per minute. The HPLC analysis of the fungal extracts revealed the presence of many natural products which may be responsible for the antimicrobial activities they elicit. Five compounds apiginine, catechine, gallic acid, keamferol and qurcetine were identified as the major compounds (Figure 4).

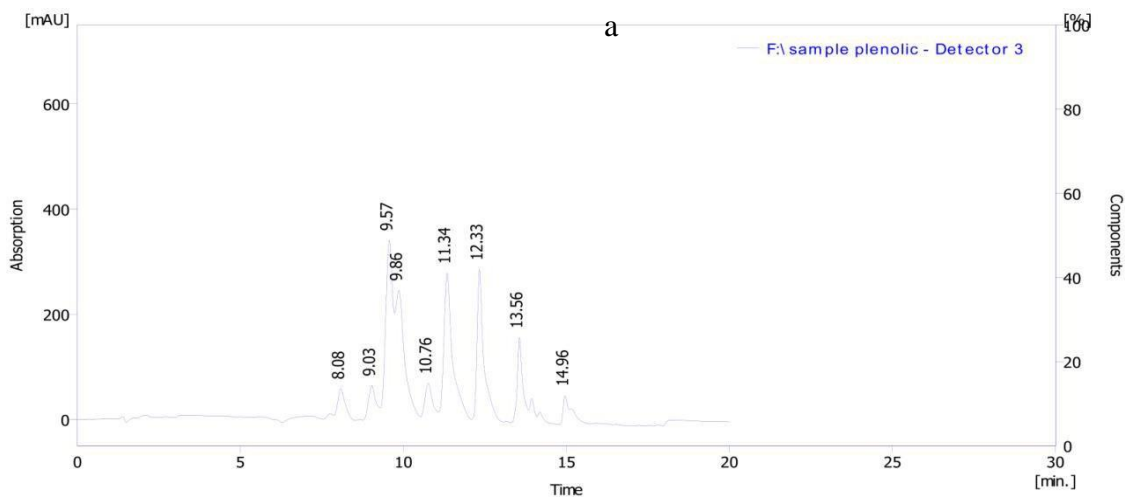

Result Table (Uncal - F:\sample plenolic - Detector 3)

| | Reten. Time [min] | Area [mAU.s] | Height [mAU] | Area [%] | Height [%] | W05 [min] | Compound Name |
|---|---|---|---|---|---|---|---|
| 1 | 8.083 | 456.983 | 41.399 | 5.0 | 4.4 | 0.19 | |
| 2 | 9.030 | 531.804 | 47.041 | 5.9 | 5.0 | 0.19 | |
| 3 | 9.567 | 1483.303 | 160.110 | 16.4 | 17.1 | 0.16 | |
| 4 | 9.860 | 1046.268 | 88.662 | 11.5 | 9.5 | 0.23 | |
| 5 | 10.763 | 578.763 | 49.844 | 6.4 | 5.3 | 0.20 | |
| 6 | 11.343 | 1285.565 | 140.844 | 14.2 | 15.0 | 0.15 | |
| 7 | 12.333 | 2190.409 | 230.264 | 24.2 | 24.6 | 0.15 | |
| 8 | 13.557 | 1122.846 | 135.114 | 12.4 | 14.4 | 0.13 | |
| 9 | 14.960 | 370.931 | 43.049 | 4.1 | 4.6 | 0.14 | |
| | Total | 9066.871 | 936.327 | 100.0 | 100.0 | | |

b

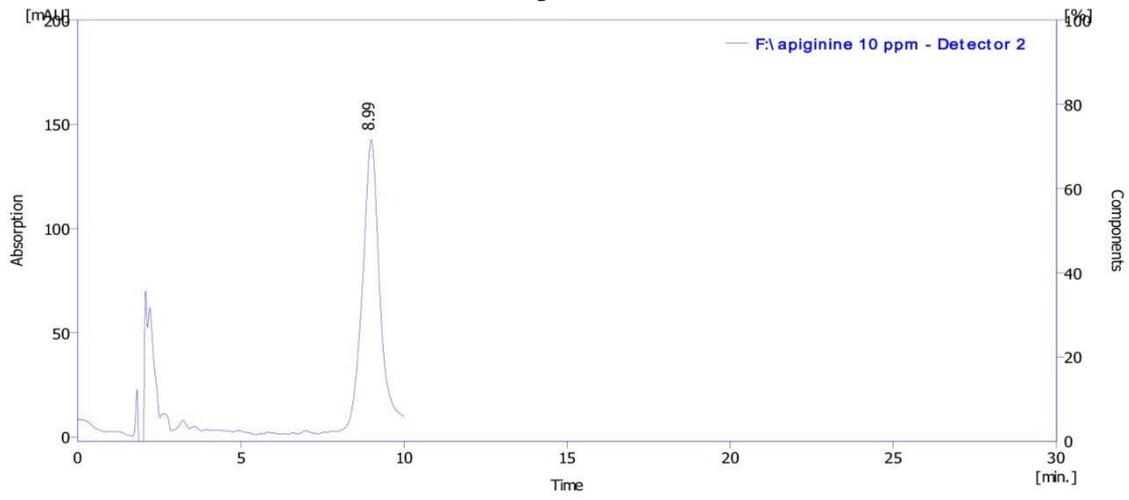

c

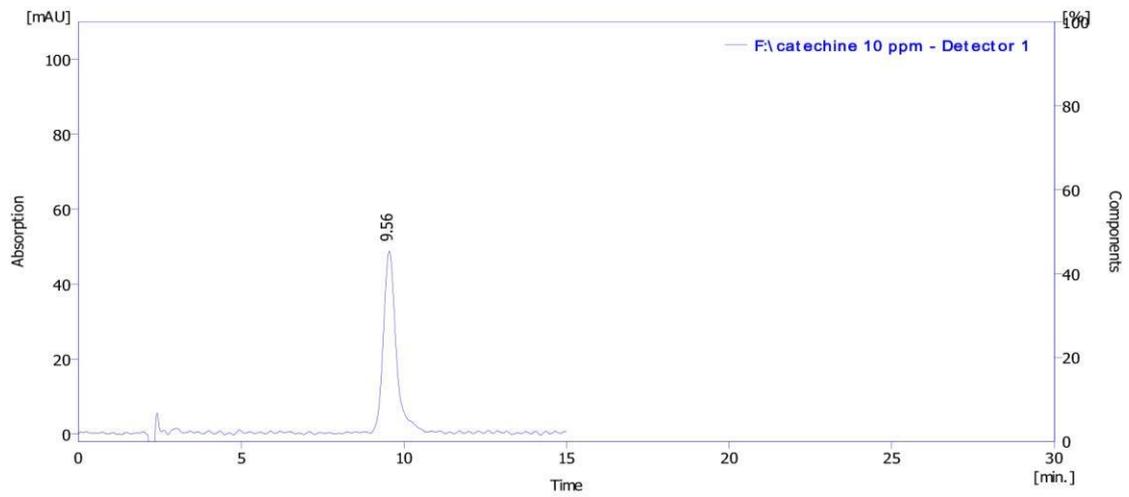

d

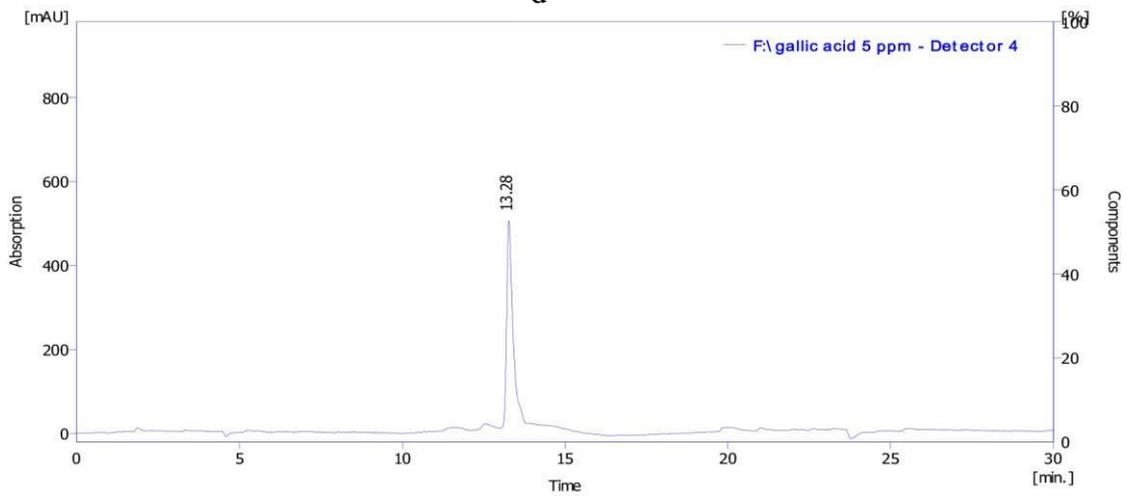

| | Reten. Time [min] | Area [mAU.s] | Height [mAU] | Area [%] | Height [%] | W05 [min] | Compound Name |
|---|---|---|---|---|---|---|---|
| 1 | 13.277 | 795.984 | 128.031 | 100.0 | 100.0 | 0.08 | |
| | Total | 795.984 | 128.031 | 100.0 | 100.0 | | |

Result Table (Uncal - F:\gallic acid 5 ppm - Detector 4)

e

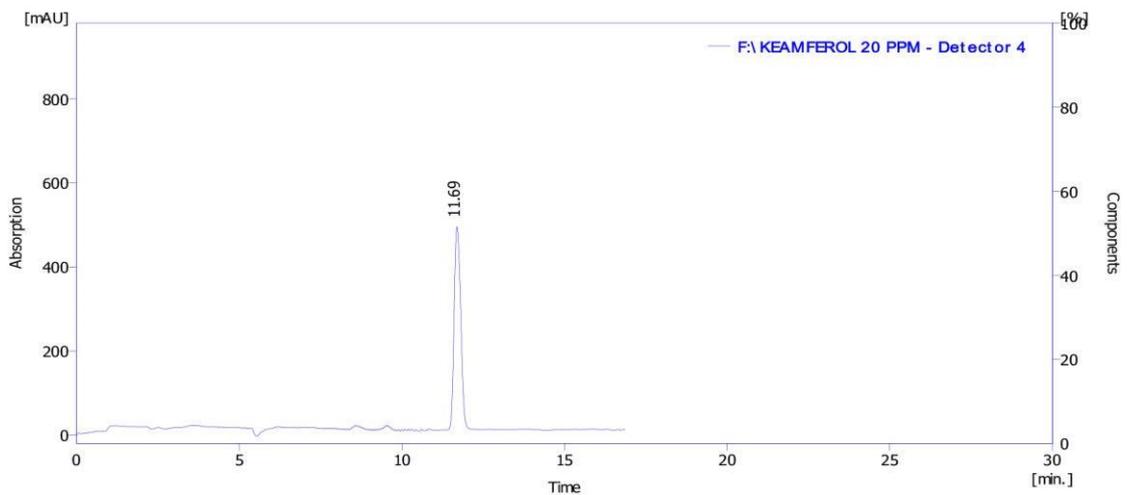

| | Reten. Time [min] | Area [mAU.s] | Height [mAU] | Area [%] | Height [%] | W05 [min] | Compound Name |
|---|---|---|---|---|---|---|---|
| 1 | 11.693 | 742.543 | 92.433 | 100.0 | 100.0 | 0.09 | |
| | Total | 742.543 | 92.433 | 100.0 | 100.0 | | |

Result Table (Uncal - F:\KEAMFEROL 20 PPM - Detector 4)

f

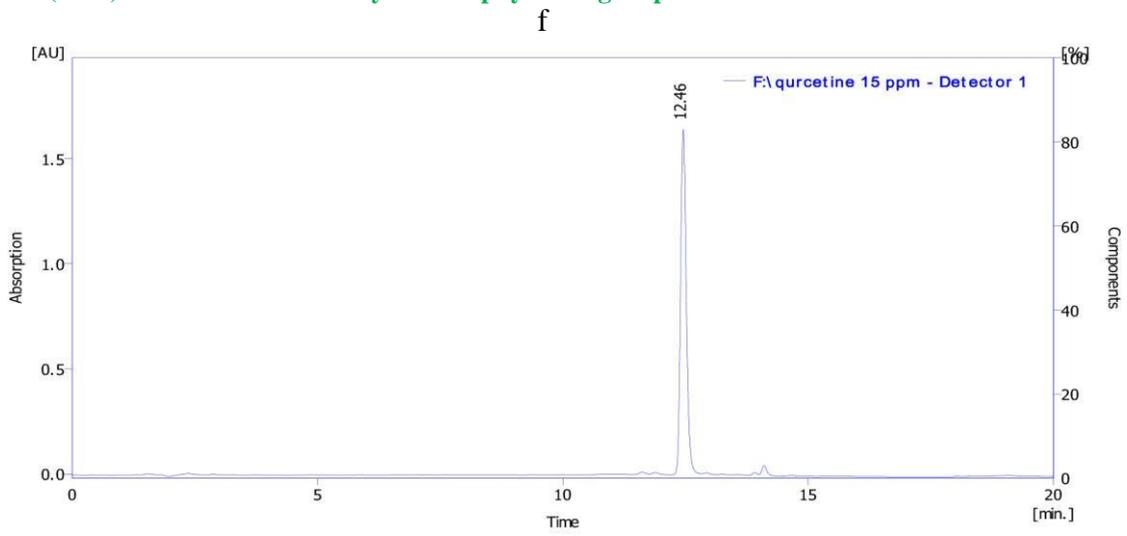

g h

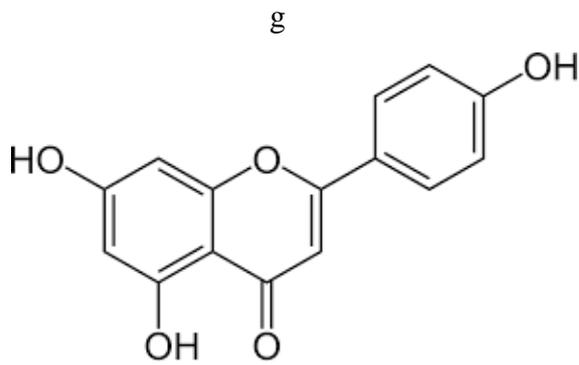

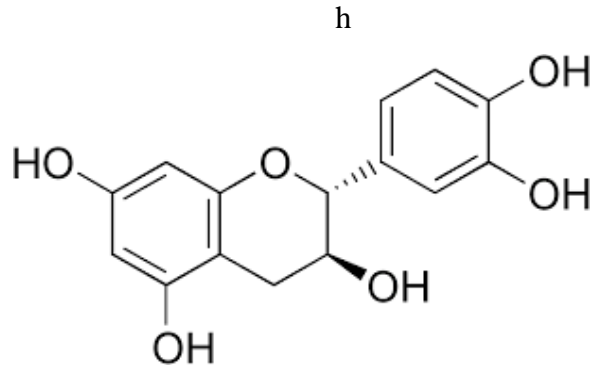

i j

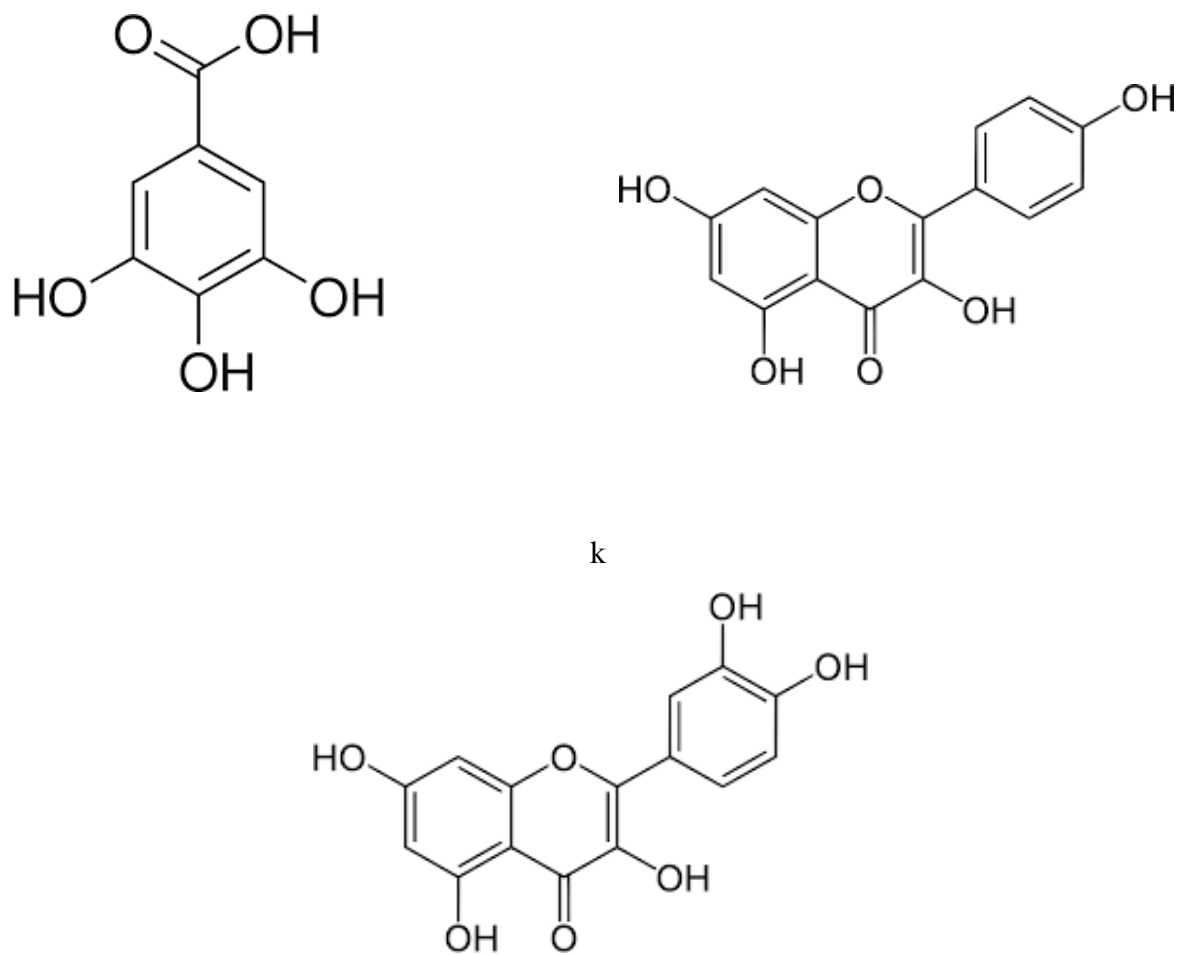

Figure 4: The HPLC data and result tables of compounds detected in *T. harzianum* extracted by ethyl acetate: (a) HPLC chromatogram showing the detection of sample plenolic; (b) apiginine; (c) catechine; (d) gallic acid; (e) kaemferol; (f) quercetin (g) Chemical structure of apiginine ($C_{15}H_{10}O_5$) https://en.wikipedia.org/wiki/Apigenin ; (h) Chemical structure of catechine ($C_{15}H_{14}O_6$) https://en.wikipedia.org/wiki/Catechin; (i) Chemical structure of gallic acid ($C_7H_6O_5$) https://en.wikipedia.org/wiki/Gallic_acid; (j) Chemical structure of kaemferol ($C_{15}H_{10}O_6$) https://en.wikipedia.org/wiki/Kaempferol; and (k) Chemical structure of quercetin ($C_{15}H_{10}O_7$) https://en.wikipedia.org/wiki/Quercetin .

**Antimicrobial activity**

Different fungal isolates were obtained from the roots of a medical plant *R. officinalis*, but the most frequency fungi selected was *T. harzianum*. Identification of this endophyte was carry out based on morphology where culture plate of was marked by the green color, also by molecular identification of the most common region which is ITS region using universal primers ITS4 and ITS5 in a standard PCR reaction. The crude extracted from *T. harzianum* roots and isolate with ethyl acetate was screened for its antimicrobial potential. This extract showed significant antimicrobial activity against
*P. aeruginosa*, *S. aureus*, , *K. pneumonia*, *B. subtilis*, and *E. coli*. The antimicrobial activity was highest against *P. aeruginosa* (35mm) at a concentration of 40μg/ml, followed by *S. aureus* (25mm) and *K. pneumonia* (25mm) at the same concentration (Table 1). The second- highest antimicrobial activity was at a concentration of 20 μg/ml against *P. aeruginosa* (25mm), followed by *K. pneumonia* (20mm) and *B. subtilis* (20mm). The lowest antimicrobial activity was against *E. coli* at all concentrations used in this study (Table 1). These antimicrobial activities might be due to some bioactive secondary metabolites founded in Entophytic fungi *T. harzianum* such as apiginine, catechine, gallic acid, kaemferol and quercetin. These results are in agreement with (Schulz et al, 2002) that several root endophytic fungi synthesis antimicrobial metabolites, also these findings are in corroborate with the previous work that the endophytes have shown the presence of phenolic compounds and flavonoids (Ramesha and Srinivas, 2014) which are known to have strong antimicrobial activities.

| S. No | Microorganisms | Distilled water (-ve control) | Gentamycin 40µg/ml ( +ve control) | 20µg/ml | 40µg/ml | 60µg/ml |
|---|---|---|---|---|---|---|
| 1. | *Pseudomonas aeuroginosa* | - | 23 | 25 | 35 | 12 |
| 2. | *Staphylococcus aureus* | - | 15 | 19 | 25 | 10 |
| 3. | *Klebsiella pneumonia* | - | 17 | 20 | 25 | 15 |
| 4. | *Bacillus subtilis* | - | 22 | 20 | 22 | 16 |
| 5. | *Escherichia coli* | - | 14 | 18 | 20 | 11 |

**Table 1:** Antibacterial activity of phenolic extract by ethyl acetate of *Trichoderma harizanum*. Zone of inhibition (Diameter in mm).

DISCUSSION

There is no doubt that Endophytic fungi are important microorganisms that are widespread in different environments, which is a source of secondary metabolites. It possesses effectiveness against many human pathogens (Desale and Bodhankar., 2013, Ramesha and Srinivas, 2014., Handayani et al., 2017). Plant roots are colonized by endophytic fungi (Girlanda and Luppi, 2006). These organisms do not harm the host, but have many benefits to the host such as improving host growth, supply the host with nutrients, etc. (Stone et al., 2000, Schulz and Boyle, 2005). Some endophytic fungi produce metabolites compounds that active biologically against fungi, bacteria, and herbicids (Schulz et al., 2002). The secondary metabolites contain active chemical groups such as flavonoids, phenols, alkaloids, terpenoids, steroids, quinines, etc (Tran et al., 2010, Aly et al., 2010, Muthu et al., 2011 ). In this study the *T. harzianum* isolates were obtained from *Rosmarinus officinalis* roots to test its ability to produce secondary metabolites that are effective against bacteria, and to determine the optimal concentrations of fungi activity and its production of secondary metabolism using agar disc diffusion method, as it is a rapid test method and shows whether the fungus has the ability to produce effective compounds outside the hypha that inhibits bacterial growth. The efficacy of fungal discs in inhibiting bacterial growth may be attributed to the ability of the fungus to produce effective metabolites compounds outside cells which inhibit the growth of a wide range of bacteria and other microorganisms (Kaul et al., 2012, Jain and Pundir., 2011). The difference in the effectiveness of the fungal species may be due to the difference in the secreted substances and their vital effect, or it may be due to the different components of the extract, such as mono and twin turbines, as well as different phenolic compounds, because different compounds have different diffusion rates which may be responsible for the differences in the inhibitory areas obtained during testing the effectiveness of the materials against sensitive organisms (Irobi et al., 1996). Here, many human pathogenic bacteria were used such as *P. aeruginosa, S. aureus*, *Klebsiella pneumoniae, Bacillus subtilis* and *Escherichia coli* which considered a dangerous species, and as a result of the mutations occurrence continuously; they became resistant to antibiotics. Therefore, it is necessary to constantly search for new types of antibiotics that are more effective against pathogenic bacteria, Endophytic fungi are among the most important sources of antibiotics (Mekawey, 2010, Fischbach, 2009). In order to obtain secondary metabolites in abundant quantities, the fungus was grown in a liquid food medium (PDB) at 25 ° C and using a vibrator for a 10-day incubation period, which is the ideal conditions for fungi growth (Ritchie et al ., 2009, Bhattacharyya and Jha., 2011), as it is known that the process of producing secondary metabolites occurs during the final stages of growth (Calvo et al., 2002), therefore the fungal isolates were extracted after 10 days. The current results are also consistent with previous studies on other fungal groups (Rosa et al., 2003, Peláez et al., 1998). The process of extracting secondary metabolites is an important stage in obtaining bioactive compounds, when extracting the crude compounds from the filtrate of fungal isolates; the appropriate solvent must be chosen depending to the nature of the extracted substance. Here, ethyl acetate was used in the extraction process, as it mixes with water in a low percentage by 33%, in addition to its low boiling point about 77 °C, therefore the process of removing it by evaporation is easy (El-Naggar., 2001). The extraction process depended on reducing the pH of the filtrate of fungal isolates to pH = 3, to convert the active substance (antagonist) extracted into ionic form, thus reducing its solubility in the aqueous solution and facilitating its extraction by ethyl acetate (Boon., 1988, Endo et al., 1986). In the present study, HPLC analysis of ethyl acetate extracts of *T. harzianum* extract reveald the presence of phenolic compounds (Table 4). Our results are consistent with those described by Ramesha and Srinivas, (2014) that the endophytes have phenolic compounds and flavonoids with displaying antimicrobial activities against pathogenic bacteria. This result is also consistent with (Desale and Bodhankar, 2013, Nwakanma et al., 2016, Tong et al., 2018, Mekawey, 2010, Fischbach, 2009 ) which confirmed that most fungal extracts have a inhibitory effect on bacteria and that fungi are an important source of antibiotics. The difference in the effectiveness of extracted compounds against bacteria may be due to several factors, including the differences in cellular structure of bacteria, as well as some of inhibitory substances are working on a specific site which is the target site. For example, many bacterial inhibitors are inhibiting the important steps of peptidoglycan production, which is a main component of the cell wall (Ghannoum and Rice., 1999). However, *E. coli* bacteria showed resistance against the secondary metabolites at all concentrations used, this may be due to the difference in structure of the cell wall between the Gram-negative and Gram-positive bacteria, as the wall of negative bacteria is characterized by low permeability to metabolites due to the presence of the outer layer, which prevents the access of the antibiotics to the target area (Yoshimura and Nikaido., 1982), also the Gram-negative bacteria wall contains lipopolysaccharides, lipoprotein and phospholipids compounds which maybe making it less permeable and resistant to antibiotics and other compounds (Wu., 2014).

**CONCLUSION**

The current study revealed that the endophytic fungi *T. harzianum* isolated from the roots of a medical plant *R. officinalis* is effective alternative sources of antimicrobial drugs, with a diversified chemical composition and vital effectiveness against other microorganisms.